# Recent results from the ANTARES deep sea neutrino telescope

Paschal Coyle on behalf of the ANTARES Collaboration

*Centre de Physiques des Particules de Marseille (CPPM), 163 Avenue de Luminy, Cedex 09 Marseille, 13280, France*

**Abstract**

The ANTARES deep sea neutrino telescope has acquired over four years of high quality data. This data has been used to measure the oscillation parameters of atmospheric neutrinos and also to search for neutrinos of a non-terrestrial origin. Competitive upper limits on the fluxes of neutrinos from dark matter annihilation in the Sun, a variety of Galactic and extra-galactic sources, both steady and transient, are presented.

## 1. Introduction

Despite a century of experimental and theoretical efforts, the exact nature of the high energy cosmic rays that bombard our Earth, their astrophysical origin and the physical processes that can accelerate hadronic particles to such extreme energies remain unclear.

As neutrinos are only produced in decays of charged hadrons they offer an unambiguous signature for those astrophysical sources in which this hadronic acceleration is present. In contrast, to more conventional probes such as protons and gamma rays, neutrinos only interact via the weak interaction. As a consequence, their observation horizon is not limited by interaction with the various ambient photon fields in the Universe. In addition, being neutral, they are not deviated by magnetic fields during their trajectory to Earth. In principle then, neutrino telescopes such as ANTARES, offer the potential for a unique 'hadronic vision' of the high energy Universe and the promise to elucidate many of the long standing questions of cosmic ray physics.

In the following, some of the various searches pursued by the ANTARES Collaboration to detect non-terrestrial neutrinos are presented.

## 2. The ANTARES neutrino telescope

Due to their small interaction cross section, neutrinos are notoriously difficult to detect. Consequently, very large volume detectors are required. Deep sea and deep ice locations provide the necessary target mass at an affordable price. Being a transparent medium, Cherenkov photons induced by the passage of relativistic muons resulting from neutrino interactions in the vicinity of the telescope are detected by a large array of photosensors. By optimising the detector design to track upcoming muons, the background from downgoing atmospheric muons is reduced. However, an irreducible background of upgoing neutrinos from cosmic ray induced atmospheric showers remains.

The ANTARES neutrino telescope [1] is located at a depth of 2475m in the Mediterranean Sea, 42km off the coast of the South of France, near Toulon. The detector comprises 885 photomultipliers distributed in a three dimension array on twelve 450m high vertical detection lines (Fig. 1). The lines are separated with a typical interline spacing of 60-70m and each line comprises 25 storeys, separated by 14.5m. A storey hosts a triplet of Optical Modules

(OMs) orientated at 45° in order to maximise the sensitivity to Cherenkov light from upcoming neutrinos. The OMs contain a 10 inch PMT protected in a 17 inch pressure resistant glass sphere. The lines are connected to a junction box, via interlink cables on the seafloor. It provides electrical power and gathers together the optical fibres from each line into a single electro-mechanical fibre optic cable for transmission of the data to and from the shore station.

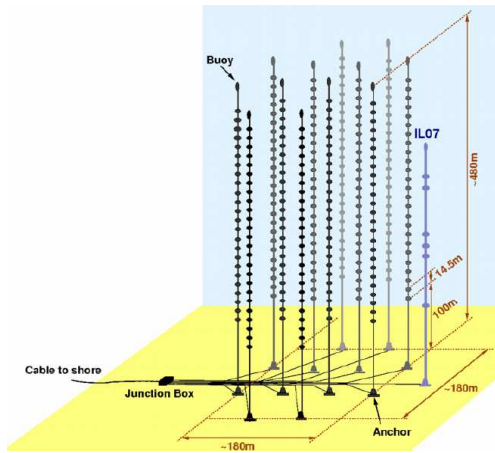

Fig. 1. The ANTARES detector configuration.

The detection lines are fixed on the seabed by a heavy weight and held taut by a buoy at the top. The lines are not rigid, but move in reaction to variations in the sea current intensity and direction. The movement of the lines is measured by an acoustic positioning system which provides a continuous and precise location (few cms) of the photosensor elements in real-time.

The infrastructure also hosts a thirteenth line, the instrumentation line (IL07), which provides measurement of environmental parameters such as sea current, temperature and also hosts a part of the AMADEUS system [2]; a test bed for the acoustic detection of ultra-high energy neutrinos. In December 2010, a secondary junction box, dedicated to host sensors for various Earth and Marine science projects, was also connected to the main junction box.

The first ANTARES detection lines were deployed in 2006 and the detector was completed in 2008. Data taking commenced during the construction phase and has been essentially continuous ever since.

The northern hemisphere location of ANTARES, provides access to the Southern sky and thus an excellent view of the Galactic plane and the Galactic centre. Together ANTARES and IceCube [3], located at the South Pole, provide continuous coverage of the full neutrino sky.

## 3. Surface Array

A good knowledge of the absolute pointing of the telescope is mandatory for astronomical studies. For ANTARES, the pointing relies on the acoustic positioning of the detection lines with respect to a surface boat located via GPS. In the absence of the detection of a bright point source of cosmic neutrinos, a method to verify the absolute pointing of the detector with respect to astrophysical coordinates is desirable. One approach is to use the Moon shadow, but as ANTARES is optimised for upgoing tracks, its sensitivity to the Moon shadow using downgoing muons is limited; a preliminary study [4] demonstrated a 2.7 σ deficit of tracks in the direction of the Moon, consistent with expectations from simulation. Although encouraging, a significant constraint on the absolute pointing would require many additional years of data.

An alternative approach has been developed in which a surface array of charged particle detectors was deployed on a surface boat and angular correlations performed between the same downgoing shower measured by the surface array and by the ANTARES detector below (Fig. 2).

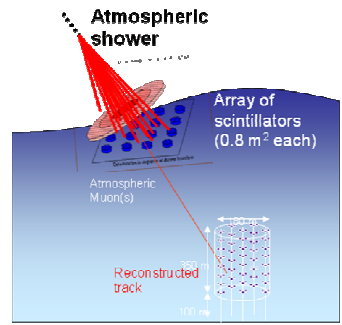

Fig. 2. The principle of the ANTARES Surface Array project.

A five day campaign was performed in October 2011, during which the boat circled around the ANTARES telescope at various radii from its centre. Applying a 10μs coincidence time window between the tracks observed in the surface array and the corresponding track observed in ANTARES, selected a total of 127 events. The derived constraints on the absolute pointing are -0.5±0.5° in zenith and 0.9±2.2° in azimuth. A second campaign is planned in the autumn of 2012 to improve these constraints, in particular that on the azimuth.

**Neutrino Oscillations**

The ANTARES telescope is optimised for the detection of neutrinos of ~TeV energies. Nevertheless, muon tracks with energies as low as 20 GeV can be reconstructed. This opens the opportunity for a measurement of neutrino oscillations using atmospheric neutrinos [5].

In the two flavour framework for neutrino oscillations, the transition probability of muon neutrinos is given by

$$P(\nu_\mu \rightarrow \nu_\mu) = 1 - \sin^2 2\theta_{23} \sin^2 (1.27 \Delta m^2_{32} L/E_\nu)$$
$$= 1 - \sin^2 2\theta_{23} \sin^2 (16200 \Delta m^2_{32} \cos\Theta/E_\nu)$$

where $\Delta m_{32}$ is the absolute difference of the squares of the mass eigenstates and $\sin^2 2\theta_{23}$ the corresponding mixing angle. For upgoing neutrinos the travel distance L is, to a good approximation, related to the zenith angle, $\Theta$, by $L = D \cdot \cos\Theta$ where D is the Earth diameter. Taking $\Delta m^2_{32} = 2.43 \cdot 10^{-3}$ eV$^2$ and $\sin^2 2\theta_{23} = 1$, the first oscillation maximum i.e. $P(\nu_\mu \rightarrow \nu_\mu) = 0$ for vertical upgoing neutrinos is at $E_\nu = 24$ GeV. Muons induced by a 24 GeV neutrino can travel up to 120m in sea water.

As the oscillation probability depends on $E_\nu/\cos\Theta$, a natural variable to study is the ratio between the reconstructed neutrino energy, $E_R$, and the reconstructed zenith angle, $\Theta_R$. The zenith angle is estimated from the muon track fit and the neutrino energy is estimated from the observed muon range in the detector. Fig. 3 shows the measured $E_R/\cos\Theta_R$ ratio for the event sample selected from the 2007-2010 data. A deficit in the low energy region is clearly visible.

A fit to the $E_R/\cos\Theta_R$ distribution yields central values of $\Delta m^2_{32} = 3.1 \cdot 10^{-3}$ eV$^2$ and $\sin^2 2\theta_{23} = 1$ with a $\chi^2$/NDF of 17.1/21; this fit is shown as the red curve in Fig. 3. The non-oscillation hypothesis yields $\chi^2$/NDF of 40/24, which has a probability of only 2.1%. The corresponding contour curve in $\Delta m_{32}$ vs. $\sin^2 2\theta_{23}$ plane is shown in Fig. 4. The results are in agreement with measurements from other experiments. This is the first such measurement of oscillation parameters by a very large volume neutrino telescope and illustrates the good understanding of the detector performance at its lowest accessible energies.

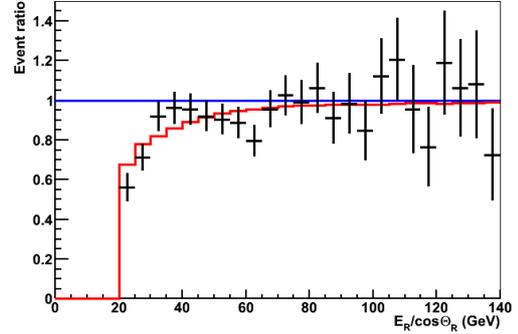

Fig. 3. Distribution of the $E_R/\cos\Theta_R$ ratio, relative to the non oscillation hypothesis. The red curve shows the result of the best fit to the oscillations parameters.

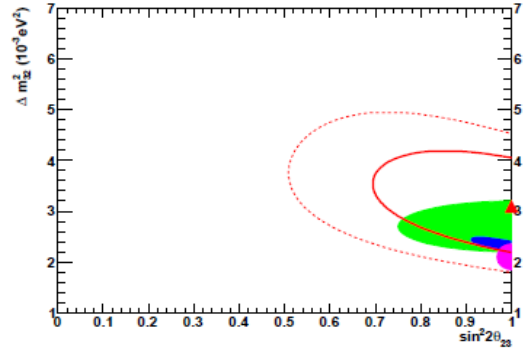

Fig. 4: The 68% and 90% C.L. contours (solid and dashed red lines) of the neutrino oscillation parameters as derived from the fit of the $E_R/\cos\Theta_R$ distribution. The best fit point is indicated by the triangle. The solid filled regions show results at 68% C.L. from K2K [6] (green), MINOS [7] (blue) and Super-Kamiokande [8] (magenta) for comparison.

## 4. Dark Matter Searches

Also in the energy region below a TeV, neutrino telescopes are sensitive to neutrinos produced by annihilation of dark matter. In many theoretical models a Weakly Interacting Massive Particle (WIMP), a relic from the Big Bang, is proposed to explain the formation of structure in the Universe and the discrepancy observed between the measured rotation curves of stars and the associated visible matter distribution in galaxies. A generic property of such WIMPs is that they gravitationally accumulate at the centre of massive bodies such as the Sun or the Earth, where they can self annihilate into normal matter. Only neutrinos, resulting from the decay of this matter, can escape from the body and be detected by a neutrino telescope.

ANTARES has performed a preliminary search for a neutrino signal originating from the direction of the Sun using the 2007-2008 data. The selection cuts are optimised to yield the best model rejection factor, based on the flux spectra provided by the WIMPSIM [9] simulation. This generator calculates the annihilation rate of WIMPs in the Sun taking into account their propagation through the Sun volume and three flavour neutrino oscillations along the trajectory to Earth. Decay channels leading to both hard ($W^+W^-$, $\tau^+\tau^-$) and soft ($b\bar{b}$) flux spectra are considered. The expected background is estimated from the data by scrambling the direction of the observed neutrino candidates. The optimised search cone around the Sun direction varies between 8°-4° depending on the assumed WIMP mass.

No significant excess of events in the direction of the Sun is observed; therefore limits on the flux of muons from the Sun shown in Fig. 5 are derived.

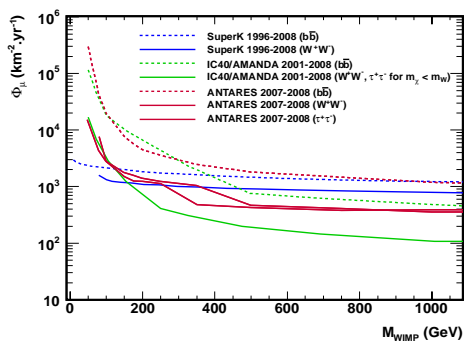

Fig. 5: Limits (90%C.L.) on the muon flux as a function of WIMP mass and its assumed decay channel.

The limits from ANTARES are competitive with those from IceCube and SuperK. In Fig. 6 these limits are converted to limits on the spin-dependent cross-section of the WIMP on protons. This choice facilitates the comparison of the indirect constraints from neutrino telescopes with those from the direct detection experiments. Also shown is a scan of supersymmetric (CMSSM) models that are compatible with the experimental constraints from the direct detection and accelerator experiments. The neutrino telescopes are able to exclude many additional models, in particular those within the focus point region. By incorporating the additional data already available and improving the event selection, a factor 3-5 improvement in sensitivity is anticipated in the next update of this analysis.

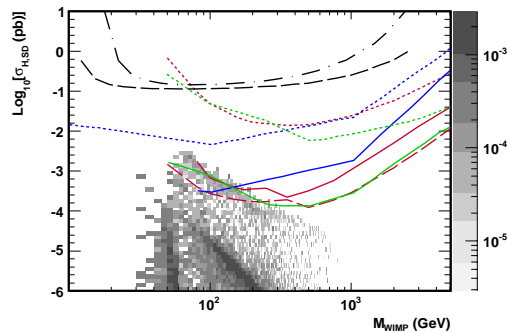

Fig. 6: Limits on the spin-dependent cross-section of WIMPS on protons inside the Sun. Non-excluded CMSSM models from a scan over its parameter space are indicated. Same legend as Fig. 5 with the addition of the direct detection limits from KIMS 2007 (dashed-dotted black line) and COUP 2011 (dashed black line).

## 5. Point Source Searches

To search for point sources of cosmic neutrinos a skymap of the arrival directions of neutrino candidates is constructed and a search performed for a localised cluster of events in excess of the uniform atmospheric backgrounds. These backgrounds are reduced to a manageable level by cuts on, the track reconstruction quality, the event-by-event estimation of the angular uncertainty provided by the track fit and finally a requirement that the track be reconstructed as upgoing. Fig. 7 shows the comparison of the distribution of the zenith angle with the corresponding Monte Carlo background simulations, a good agreement is observed.

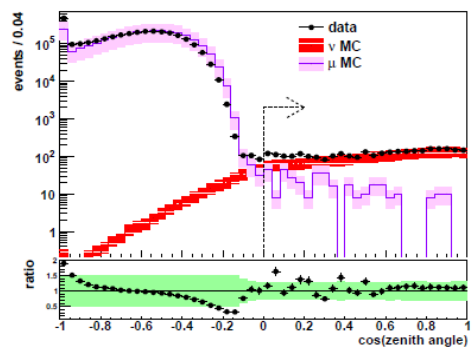

Fig. 7: The reconstructed zenith angle of the selected events compared to the Monte Carlo expectations for the atmospheric neutrino and muon background. The upgoing events (indicated by the arrow) are selected for inclusion in the skymap.

After these quality cuts, the median uncertainty on the reconstructed neutrino direction, assuming an $E^{-2}$ neutrino energy spectrum, is 0.5±0.1 degrees.

Applied to the 2007-2010 data recorded by ANTARES (corresponding to a total livetime of 813 days), a total of 3058 events pass the optimised selection criteria [10]. Their reconstructed arrival directions are shown in Fig. 8. An unbinned maximum likelihood method, based on the estimated point spread function and incorporating an energy estimator based on the number of observed hits, was used to search for clusters of events anywhere in the skymap, consistent with a point-like source emitting with an $E^{-2}$ neutrino flux spectra.

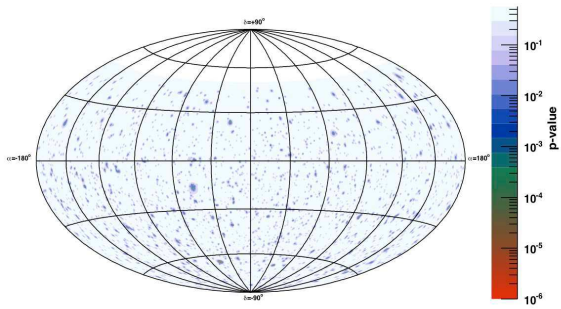

Fig. 8: Sky map in equatorial coordinates showing the p-values obtained for the point-like clusters evaluated in the full-sky scan; the penalty factor accounting for the number of trials is not considered in this calculation.

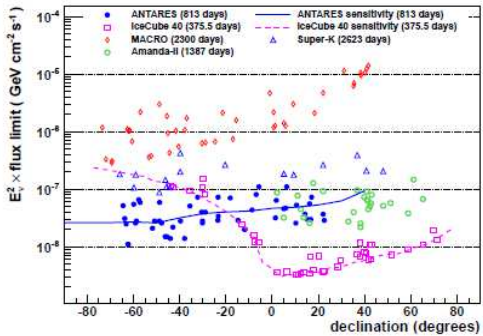

Fig. 9: Upper limits (at 90% CL) on the $E^{-2}$ neutrino flux from selected candidate sources as function of their declination. Upper limits from other experiments are also indicated. The dotted lines show the expected median sensitivities.

The most significant cluster of events was found at $(\alpha,\delta)=(-46.5°,-65.0°)$. Pseudo-experiments taking into account systematic uncertainties of the angular resolution and the acceptance of the detector were used to determine a post-trial significance of 2.6%. An additional 'candidate search' in the direction of 51 a-priori selected candidate sources, i.e. interesting objects observed in high energy gamma rays, also did not yield any significant excess. Upper limits on the neutrino flux from these sources were derived as shown in Fig. 9; also indicated is the median expected sensitivity of the analysis.

The ANTARES limits represent the best limits for point sources in the Southern sky. It should be noted that for the IceCube limits in the Southern sky an energy threshold above 1 PeV is applied in order to mitigate the downgoing muon background.

## 6. Searches for Extended Sources

### 6.1. Supernova remnant RXJ1713.7-3946

The sensitivity of the ANTARES point source search to spatially extended sources such as the shell-type supernova remnant RX J1713.7-3946, a well motivated potential site of hadronic acceleration, was also evaluated [10]. Assuming the neutrino flux to be related to the gamma flux observed by HESS and taking into account its measured extension, the experimental limit is a factor 8.8 than the theoretical prediction; the most restrictive available for the emission model considered.

### 6.2. Fermi Bubbles

The so called 'Fermi Bubbles' (FB) are an excess of gamma ray emission observed in enormous (~10kpc) bilateral bubbles above and below the Galactic plane. It has been proposed [11] they are due to a population of relic cosmic ray protons and heavier ions injected by processes associated with extremely long timescale (>8 Gyr) and high density star-formation in the Galactic center. If this case, they are also a potential site for high energy neutrino production.

A dedicated search for an excess of neutrinos in the region of the FBs has been performed by comparing the rate of high energy events observed in the region of the FBs (ON zone) to that observed in equivalent areas of the Galaxy excluding the FBs (OFF zones). Fig. 10 shows the distribution of the number of hits (a proxy for an energy measurement) for the selected events in the ON and OFF regions. As no excess of events is observed in the ON region above the chosen energy cut, limits are placed on possible fluxes of neutrinos for various assumptions on the energy cutoff at the source (Fig. 11).

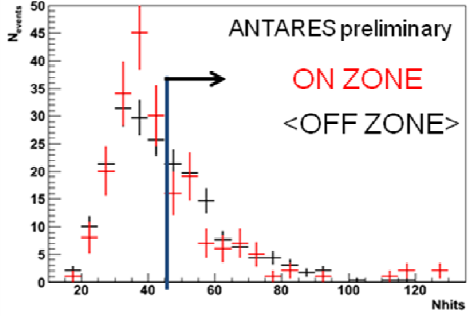

Fig. 10: Distribution of the number of hits for events inside (ON zone) and outside (OFF zone) the FB regions. The 'energy' cut is indicated by the arrow.

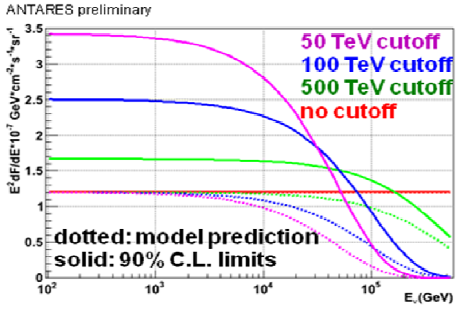

Fig. 11: Upper limits (at 90% C.L.) on the $E^{-2}$ neutrino flux from the Fermi Bubbles as a function of the assumed energy cutoff. The corresponding model predictions are indicated by the dotted curves.

## 7. Searches for Transient Sources

As transient sources emit only for a limited time period, the atmospheric backgrounds can be significantly reduced by searching in a time window coincidence with the duration of the flaring event. The length of time window may be relatively short (seconds→minutes) for the case of Gamma Ray Bursts or longer (days→months) for the case of flaring blazars or micro quasars.

An alternative approach, also pursued by ANTARES, is to generate online neutrino alerts for 'special' neutrino events [12]. These alerts are then used to trigger follow up by a network of optical telescopes pointed in the direction of the neutrino. In principle, this strategy allows a possible detection of GRBs/supernovae that might otherwise have been missed by the usual monitoring observatories.

### 7.1. Flaring Microquasars

Microquasars are Galactic binary systems comprising a compact object accreting mass from a companion star. They display relativistic ejections of matter in the form of jets, which could lead to the production of neutrinos when the jets interact with the surrounding matter.

The ANTARES analysis identified six microquasars with x-ray or gamma-ray outbursts in the 2007-2010 satellite data (RXTE/ASM, Swift/BAT and Fermi/LAT). No significant excess of neutrino events in spatial and time coincidence with the flares was found. The obtained limits on the neutrino flux are shown in Fig. 12 and compared with predicted fluxes based on the model of Ref. [13] modified to incorporate an energy cutoff of 100 TeV at the source. It can be seen that the predicted fluxes from GX339-4 and CygX-3 are close to the experimental limits.

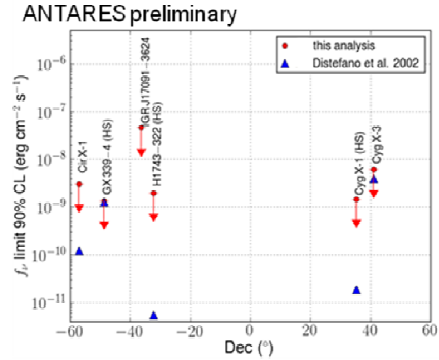

Fig. 12: Upper limits (at 90% C.L.) on the $E^{-2}$ neutrino flux from a variety of microquasars as function of their declination.

### 7.2. Flaring Blazars

For this analysis [14], a total of ten blazars demonstrating periods (days/weeks) of enhanced emission have been selected based on a study of the Fermi-LAT light curves ($E_\gamma$>100 GeV). For one of these AGNs (3C279), a single neutrino candidate was found to be in spatial (0.56°) and time coincidence with the flare. The post-trial probability for this to occur randomly was evaluated as 10%. The 90% C.L. upper limits derived on the neutrino fluence for these AGNs are in the range 1-23 GeV.cm$^{-2}$.

*7.3. Sources emitting Gravitational Waves*

The gravitational wave detectors VIRGO and LIGO are designed to detect GW emission from the same types of transient sources that could emit neutrinos. In 2007, LIGO (S5) and VIRGO (VSR1) operated in conjunction for a science run. Using the ANTARES neutrino events as a directional trigger for the GW search reduces the trial factor associated with a full sky GW search, thereby augmenting the sensitivity for potential sources that emit GWs and neutrinos simultaneously. An analysis [15] of the 2007 ANTARES observed no statistical significant correlation. Limits on the distance of occurrence of NS-BH and NS-NS mergers of around 10 Mpc were derived.

*7.4. Gamma Ray Bursts: Optical Follow Up*

The ANTARES online data acquisition system has the capability to rapidly filter and reconstruct events in real-time. Since 2009, neutrino alerts have been regularly sent, typically once per month, to a network of fast-response, wide field of view (1.9°x1.9°), robotic telescopes (TAROT, ROTSE, ZADKO). The neutrino alerts are produced either for neutrino doublets occurring within 15 minutes and 3° of each other or single neutrino events having an estimated energy above about 5 TeV. The rapid response of the system (~20s) is particularly well suited for the detection of optical afterglows from GRBs, while a longer term observation strategy over several weeks is suitable for the detection of slower transients such as core-collapse supernovae explosions.

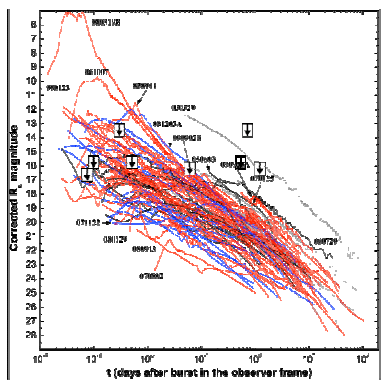

Fig. 13: The constraint on the magnitude of optical emission for a GRB in the direction of the neutrino alert. Also shown are typical time dependence for a selection of GRB afterglows.

For the 2009 data, no optical counterpart associated with a neutrino alert was observed. Fig. 13 shows the limits on the magnitude of a possible GRB afterglow associated to the observed neutrinos. Also shown is a selection of time profiles of typical GRB afterglows; for those alerts having a rapid response, the system has the requisite sensitivity to detect a typical GRB afterglow, if the neutrino really did originate from a GRB.

## 8. Summary

ANTARES is the first and only deep sea neutrino telescope currently in operation. It has been taking data essentially continuously and smoothly since the first lines were deployed in 2007. A variety of competitive searches are underway in an effort to detect the first sources of high-energy cosmic neutrinos. The adoption of a multi-messenger approach significantly enhances the sensitivity of these searches.

Although not addressed in this proceedings, the ANTARES infrastructure is also a unique multi-disciplinary deep sea observatory that has catalysed a number of innovative and synergetic studies in the fields of Earth and Marine sciences [16].

The successful demonstration by ANTARES of the deep sea technique for neutrino astronomy is an important step towards the European-wide effort to construct the future multi-kilometre cube observatory, KM3NeT, in the Mediterranean Sea [17].